\documentclass{emulateapj}
\usepackage{amsmath}     
\usepackage{wasysym}           
\usepackage{graphicx}
\usepackage{amssymb}
\usepackage{epstopdf}
\usepackage{mathrsfs}
\usepackage[T1]{fontenc}
\usepackage{natbib}

\begin{document}
\title{Viscous boundary layers of radiation-dominated, relativistic jets. I. The two-stream model}
\author{Eric R. Coughlin\altaffilmark{1} and Mitchell C. Begelman\altaffilmark{1}}
\affil{JILA, University of Colorado and National Institute of Standards and Technology, 440 UCB, Boulder, CO 80309}
\email{eric.coughlin@colorado.edu, mitch@jila.colorado.edu}
\altaffiltext{1}{Department of Astrophysical and Planetary Sciences, University of Colorado, UCB 391, Boulder, CO 80309}

\begin{abstract}
Using the relativistic equations of radiation hydrodynamics in the viscous limit, we analyze the boundary layers that develop between radiation-dominated jets and their environments. In this paper we present the solution for the {}{self-similar, 2-D, plane-parallel} two-stream problem, wherein the jet and the ambient medium are considered to be separate, interacting fluids, and we compare our results to those of previous authors. (In a companion paper we investigate an alternative scenario, known as the free-streaming jet model.) Consistent with past findings, we show that the boundary layer that develops between the jet and its surroundings creates a region of low-density material. {}{These models may be applicable to sources such as super-Eddington tidal disruption events and long gamma-ray bursts.}
\end{abstract}


\section{Introduction}
Astrophysical jets almost certainly exist as aggregates of massive particles, magnetic fields, and radiation. In certain scenarios, however, the contribution of radiation to the energetics of the outflow far outweighs those of the particles and magnetic fields, meaning that one can essentially neglect the presence of the latter two entities.

One such scenario occurs in the collapsar model of long gamma-ray bursts (GRBs; \citealt{woo93, mac99}). In this model, the core of a massive, evolved star collapses directly (or with a short-lived neutron star phase) to a black hole during the infall stage of a type-II supernova. The energy released by the material accreting onto the {}{black} hole, and ultimately observed as the gamma-ray emission, is collimated into bipolar jets -- the jet formation confirmed by energetics arguments \citep{wax98, fru99, fra01} and the observations of breaks in the X-ray afterglow light curves \citep{pan07, dad08, rac09} -- and is often sufficient to unbind the stellar envelope, resulting in a supernova (\citealt{gal98, ber04, kam09, lev14}, but see \citealt{fyn06}). If one assumes that the mass of the remnant {}{black} hole is on the order of a few solar masses, its accretion luminosity exceeds the Eddington limit by roughly ten orders of magnitude, meaning that radiation pressure, even if the flux is nearly isotropic, is likely an important mechanism for driving and sustaining the outflow (the fireball model; \citealt{ree92}). Even if the jet is launched by magnetohydrodynamical mechanisms \citep{bla77, bla82}, radiation could still play a prominent role in determining the dynamics of the jet. Arguments concerning the time necessary for the jet to break through the stellar envelope also seem to disfavor Poynting-dominated jets {}{(\citealt{bro15}; but see \citealt{mun13})}.

Jets produced during tidal disruption events (TDEs; \citealt{gia11}) -- when a star is destroyed by the tidal force of a supermassive black hole -- could provide another class of radiation-dominated outflow. After the star is tidally disrupted, roughly half of the shredded debris remains bound to the black hole and returns to the tidal disruption radius. If the black hole has a mass less than roughly $10^{7}M_{\astrosun}$, that rate of return can exceed the Eddington limit of the {}{black} hole by orders of magnitude for a significant amount of time (on the order of days to months; \citealt{eva89, str09}). Provided that this material can rapidly accrete onto the {}{black} hole, which is likely the case due to the tidal dissipation of kinetic energy \citep{koc94, gui14} and relativistic precession effects \citep{ree88, eva89}, the energy released during the accretion process will also be supercritical. It was during this supercritical phase that the event \emph{Swift} J1644+57 was seen to have an associated jetted outflow \citep{bur11, blo11, can11, zau11} (the source \emph{Swift} J2058+05 may provide another example of a jetted, super-Eddington TDE; \citealt{cen12}). Although the jet launching mechanism for this event is uncertain, the magnetic field of the tidally-disrupted star, assuming its flux is approximately conserved, is almost certainly insufficient to power the outflow. Therefore, unless one invokes the existence of a fossil magnetic field \citep{tch14, kel14}, the radiation pressure associated with the accretion luminosity likely plays some role in powering the jet. At any rate, the radiation released during the supercritical accretion process affects the dynamics of the collimated outflow and contributes substantially to its overall energy and momentum. 

Collapsar jets inject a significant amount of energy into the overlying stellar envelope as they punch their way into the circumstellar medium, creating a pressurized ``cocoon'' of shocked material with which the jet interacts \citep{mor07, laz07, lop13}, and the progenitors themselves -- usually taken to be highly-evolved, Wolf-Rayet stars \citep{mat03, woo06} -- are likely sustained primarily by radiation pressure. The initial stages of collapsar jet propagation are therefore characterized by the transfer of energy and momentum between two radiation-dominated fluids. Because of the high accretion rates and low specific angular momentum, the fallback disks generated during the super-Eddington phase of TDEs are likely optically and geometrically thick and radiation pressure-supported \citep{ree88, loe97, cou14a}. In the zero-Bernoulli accretion (ZEBRA) model of \citet{cou14a}, for example, the accreting material is inflated into a quasi-spherical envelope that surrounds the {}{black} hole. The dynamics of the jets of supercritical TDEs are therefore also modulated by the presence of a radiation pressure-supported, external medium.

Previous authors used the supersonic propagation of the outflow to model the interaction of the collapsar jet with the overlying envelope as an oblique shock-boundary layer structure \citep{bro07, koh12, koh12b, koh14}. In these models, the outflow is assumed to consist of a perfect, single fluid with a relativistic equation of state, and some have included the presence of magnetic fields (e.g., \citealt{koh12b}). A more realistic picture, however, is obtained by considering the jet as a composite of the massive scatterers present in the outflow and the radiation that accompanies it. Indeed, this approach constitutes the underlying framework of radiation hydrodynamics. Furthermore, non-ideal, i.e., viscous, effects will tend to ``smear'' the discontinuities otherwise present in the system, resulting in a more gradual transition of the fluid quantities between the jet and the environment. 

In the limit that the mean free path of a photon is small, radiation acts like an effective viscosity, with a coefficient of dynamic viscosity that depends both on the radiation pressure and the density of scatterers (see section 2 of this paper), and transfers momentum and energy between neighboring fluid elements \citep{wei71, loe92}. In the radiation-dominated interaction between collapsar jets and TDE jets and their respective ambient media, the effects of radiation viscosity should be quite large. \citet{ara92} considered the effects of radiation viscosity on the evolution of boundary layers in the two-stream approximation (see section 3 of this paper); their treatment, however, was of a non-relativistic nature, meaning that their results have limited applicability to GRB and TDE jets. 

In this paper, the first of two, we consider the effects of radiation viscosity on the propagation of relativistic jets in radiation-rich environments. In section 2 we present the equations of radiation hydrodynamics in the viscous limit. Section 3 applies those equations to the two-stream problem, wherein the jet and ambient medium are considered to be two separate, interacting fluids, and we compare our results to the non-relativistic treatment of \citet{ara92}. In section 4 we discuss the results of the analysis and {}{comment upon} the application of our models to the jets produced by supercritical TDEs such as \emph{Swift} J1644+57, GRBs{}{, and other astrophysical systems}. In a second paper \citep{cou15} we present an alternate model, the free-streaming jet solution, and compare it to the two-stream solution presented here.

\section{Governing equations}
As mentioned in the introduction, radiative forces behave like an effective viscosity in the presence of shear, when the change in fluid quantities across the mean free path of a photon is small. The precise means by which this viscous coupling manifests itself can be determined by analyzing the Boltzmann equation.

\citet{cou14b} used the general relativistic Boltzmann equation for Thomson scattering to discern the effects of radiation viscosity in the presence of both relativistic velocities and gravitational fields (i.e., in accelerating reference frames). Instead of reproducing their work here, we will simply quote the equations of radiation hydrodynamics for a cold gas (gas pressure much less than both the gas rest mass density and the radiation pressure) that result from their analysis{}{ (see their equation 49)}:

\begin{multline}
\nabla_{\mu}\bigg{[}\bigg{\{}\rho'+\frac{4}{3}e'\bigg{(}1-\frac{10}{9}\frac{1}{\rho'\kappa}\nabla_{\alpha}U^{\alpha}\bigg{)}\bigg{\}}U^{\mu}U^{\nu}\bigg{]}+\frac{1}{3}g^{\mu\nu}\partial_{\mu}e' \\ 
-\frac{8}{27}\nabla_{\mu}\bigg{[}\frac{e'}{\rho'\kappa}\Pi^{\mu\sigma}\Pi^{\nu\beta}\bigg{(}\nabla_{\sigma}U_{\beta}+\nabla_{\beta}U_{\sigma}+g_{\beta\sigma}\nabla_{\alpha}U^{\alpha}\bigg{)}\bigg{]} \\ 
-\frac{1}{3}\nabla_{\mu}\bigg{[}\frac{e'}{\rho'\kappa}\bigg{(}\Pi^{\mu\sigma}U^{\nu}+\Pi^{\nu\sigma}U^{\mu}\bigg{)}\bigg{(}4U^{\beta}\nabla_{\beta}U_{\sigma}+\partial_{\sigma}\ln{}e'\bigg{)}\bigg{]} = 0 \label{radhydroco}.
\end{multline}
Here the speed of light has been set to one, Greek indices range from 0 -- 3, $\rho'$ is the fluid rest-frame density of scatterers, $e'$ is the fluid rest-frame radiation energy density, $\kappa$ is the scattering opacity (in units of cm$^2$ g$^{-1}$), $g_{\mu\nu}$ is the metric of the spacetime, $\nabla_{\mu}$ is the covariant derivative, $U^{\mu}$ is the four-velocity of the flow, and $\Pi^{\mu\nu} = U^{\mu}U^{\nu}+g^{\mu\nu}$ is the projection tensor. The Einstein summation convention has been adopted here, meaning that repeated upper and lower indices imply summation. This equation shows that the coefficient of dynamic viscosity, $\eta$, for an optically-thick, radiation-dominated gas is

\begin{equation}
\eta = \frac{8}{27}\frac{e'}{\rho'\kappa} \label{etaeq}, 
\end{equation}
which agrees with previous findings \citep{bla85, loe92}; note that this specific coefficient is only for the case when the gas and radiation interact through Thomson scattering. The gas energy equation, which will also be useful for us, can be obtained by contracting equation \eqref{radhydroco} with $U_{\nu}$, which we can show becomes {}{(see equation 50 of \citealt{cou14b})}

\begin{multline}
\nabla_{\mu}(e'U^{\mu})+\frac{1}{3}e'\nabla_{\mu}U^{\mu} = \frac{4}{3}\frac{10}{9}\nabla_{\mu}\bigg{[}\frac{e'}{\rho'\kappa}U^{\mu}\nabla_{\alpha}U^{\alpha}\bigg{]} \\
+\frac{8}{27}\frac{e'}{\rho'\kappa}\bigg{(}\nabla_{\sigma}U_{\beta}+\nabla_{\beta}U_{\sigma}+g_{\sigma\beta}\nabla_{\alpha}U^{\alpha}\bigg{)}\Pi^{\mu\sigma}\nabla_{\mu}U^{\beta} \\
+\frac{1}{3}\Pi^{\mu\sigma}\nabla_{\mu}\bigg{[}\frac{e'}{\rho'\kappa}\bigg{(}4U^{\beta}\nabla_{\beta}U_{\sigma}+\partial_{\sigma}\ln{e'}\bigg{)}\bigg{]} \\+\frac{1}{3}\frac{e'}{\rho'\kappa}\bigg{(}4U^{\beta}\nabla_{\beta}U_{\sigma}+\partial_{\sigma}\ln{e'}\bigg{)}\bigg{(}2U^{\mu}\nabla_{\mu}U^{\sigma}+U^{\sigma}\nabla_{\mu}U^{\mu}\bigg{)}
\label{gasenergyco}.
\end{multline}
To close the system, we require that the normalization of the four-velocity be upheld and that particle flux be conserved:

\begin{equation}
U_{\mu}U^{\mu} = -1,
\end{equation}
\begin{equation}
\nabla_{\mu}\bigg{[}\rho'U^{\mu}\bigg{]} = 0 \label{masscont}.
\end{equation}
Equations \eqref{radhydroco} and \eqref{gasenergyco} -- \eqref{masscont} constitute six linearly independent equations for the six unknowns $U^{\mu}$, $e'$, and $\rho'$.

In addition to the energy density of the radiation, $e'$, one can also calculate the number density of photons by requiring that the number flux, $F^{\mu}$, be conserved. One can show \citep{cou14b} that the equation $\nabla_{\mu}F^{\mu} = 0$ becomes, in the viscous limit, 

\begin{multline}
\nabla_{\mu}\bigg{[}N'U^{\mu}\bigg{]} \\
 = \nabla_{\mu}\bigg{[}\frac{1}{\rho'\kappa}\bigg{(}\frac{10}{9}N'U^{\mu}\nabla_{\sigma}U^{\sigma}+N'U^{\alpha}\nabla_{\alpha}U^{\mu}+\frac{1}{3}\Pi^{\mu\sigma}\nabla_{\sigma}N'\bigg{)}\bigg{]} \label{fluxeq},
\end{multline}
where $N'$ is the rest-frame number density of photons. Once we solve the equations of radiation hydrodynamics for the four-velocity of the fluid and the mass density of scatterers, we can solve equation \eqref{fluxeq} for the number flux of photons. 

The goal of the next two sections is to apply equations \eqref{radhydroco} and \eqref{gasenergyco} -- \eqref{fluxeq} to the boundary layers established between fast-moving jets and their environments. For a more thorough discussion of the nature of the equations of radiation hydrodynamics in the viscous limit, we refer the reader to \citet{cou14b}.

\section{Two-stream boundary layer}
\citet{ara92} considered the \citet{bla08} boundary layer problem, wherein one analyzes the dynamics of viscous flow over a semi-infinite, rigid plate, with the viscous effects attributed to radiation. They showed, however, that the requirement that both velocity components vanish on the lower plate, the no-slip condition, results in a divergent boundary layer thickness. The authors then examined the more physical scenario of two interacting fluids, one moving at some asymptotic velocity and the other asymptotically stationary, known as the two-stream approximation. In this case the no-slip condition no longer applies, and they were able to show that the boundary layer thickness remains finite.

The treatment of \citet{ara92} was non-relativistic, meaning that their results have limited utility when one considers the boundary layers established between GRBs and super-Eddington TDEs and their ambient media. The enthalpy of the radiation, which is not ignorable in radiation-dominated flows, was also not included in their momentum equation. Here we extend their analysis to the relativistic regime and we include the radiation enthalpy.

\subsection{Basic setup}
We assume that the flow is plane-parallel with no variation in the $x$-direction. The covariant derivatives in equations \eqref{radhydroco} and \eqref{gasenergyco} -- \eqref{masscont} can therefore be replaced by ordinary partial derivatives. Even though this simplification significantly reduces their complexity, the most compact representation of the equations is still given by \eqref{radhydroco} and \eqref{gasenergyco} -- \eqref{masscont} with $\nabla_{\mu} \rightarrow \partial/\partial{x}^{\mu}$, so we do not write them again here.

The majority of the motion is along the $z$-axis, meaning $v_z \gg v_y$ and $v_x \equiv 0$. At some initial point $y = z = 0$, the ``jet,'' whose constant, asymptotic ($y \rightarrow \infty$) velocity is denoted $v_j$, encounters the ambient medium, the asymptotic ($y \rightarrow -\infty$) velocity of which is zero. The asymptotic densities of the jet and the ambient medium, denoted $\rho'_j$ and $\rho'_a$, respectively, are both taken to be constant. The line $y = 0$ represents the surface that divides the jet and ambient material, and consistent with any boundary layer analysis, we also assume that the gradient along $y$ is much greater than that along $z$, so $\partial/\partial{z} \ll \partial/\partial{y}$.

\subsection{Boundary layer equations}
The complexity of equations \eqref{radhydroco} and \eqref{gasenergyco} -- \eqref{masscont} can be reduced by introducing the boundary layer thickness $\delta{y}$ such that $\delta \sim \delta{y}/\delta{z}$ is a small parameter when $\delta{z}$ is chosen to be a typical length scale in $z$. By keeping terms only to lowest order in $\delta$, we will recover a set of reduced boundary layer equations. 

To determine how $\delta$ depends on asymptotic fluid quantities (it is the reciprocal of the square root of the Reynolds number in the classical Blasius boundary layer), we compare the lowest-order terms in $\delta$ in the gas energy equation to the inviscid terms. By equating these terms we are requiring that the viscous heating of the radiation contribute a non-negligible increase in the entropy of the fluid, but because the inviscid terms are proportional to the divergence of the four-velocity (see equation \ref{gasen1}), this equality can only be true when the gas is compressible, i.e., when the flow velocity is supersonic. When the flow becomes very subsonic, the energy equation can be ignored as the fluid is essentially incompressible. Making the substitutions $\partial/\partial{y} \sim 1/\delta{y}$, $\partial/\partial{z} \sim 1/z$, $v \sim v_j$ and $\rho' \sim \rho'_0$ {}{in equation \eqref{gasenergyco}}, where $\rho'_0 = \rho'_j$ if we are in the jet ($y>0$) or $\rho'_a$ if we are in the ambient medium ($y < 0$), we find that the boundary layer thickness scales as

\begin{equation}
\delta^2 \sim \frac{1}{\rho'_0\,\kappa\,{z}\,\Gamma_j\,v_j} \label{deltaeq},
\end{equation}
{}{where $\Gamma_j = (1-v_j^2)^{-1/2}$ is the Lorentz factor of the jet.} The boundary layer thickness therefore scales roughly as $1/\sqrt{\tau}$, where $\tau \sim \rho_0'\kappa{z}$ is the fluid-frame optical depth along $z$.

We can now use our expression for $\delta$ \eqref{deltaeq} to keep only lowest-order terms in equations \eqref{radhydroco} and \eqref{gasenergyco} -- \eqref{masscont}. The resulting $\nu = y$, $\nu = z$, gas-energy and continuity boundary layer equations are, respectively, 

\begin{equation}
\frac{\partial{e'}}{\partial{y}} = 0 \label{ymom1},
\end{equation}
\begin{equation}
\frac{\partial}{\partial{x^{\mu}}}\bigg{[}\bigg{(}\rho'+\frac{4}{3}e'\bigg{)}U^{\mu}\,\Gamma{v_z}\bigg{]}+\frac{1}{3}\frac{de'}{dz}-\frac{8}{27}\frac{\partial}{\partial{y}}\bigg{[}\frac{e'}{\rho'\kappa}\frac{\partial{}}{\partial{y}}\big{[}\Gamma{v_z}\big{]}\bigg{]} = 0 \label{zmom1},
\end{equation}
\begin{equation}
\frac{\partial}{\partial{x^{\mu}}}\bigg{[}e'U^{\mu}\bigg{]}+\frac{1}{3}e'\frac{\partial{U^{\mu}}}{\partial{x^{\mu}}} = \frac{8}{27}\frac{e'}{\rho'\kappa}\frac{\partial{U_{\mu}}}{\partial{y}}\frac{\partial{U^{\mu}}}{\partial{y}} \label{gasen1},
\end{equation}
\begin{equation}
\frac{\partial}{\partial{z}}\bigg{[}\rho'\Gamma{v_z}\bigg{]}+\frac{\partial}{\partial{y}}\bigg{[}\rho'\Gamma{v_y}\bigg{]} = 0 \label{masscont1}.
\end{equation}
The first of these demonstrates, as in the non-relativistic limit, that the pressure is constant across the boundary layer. For the remainder of this section we will assume that $e'(z) = e'_0$, i.e., that the radiation energy density is independent of $z$.

\subsection{Self-similar approximation}
The solution to the continuity equation \eqref{masscont1} can be obtained by introducing the stream function $\psi$ through the definitions

\begin{equation}
\rho'\kappa\Gamma{v_z} = \frac{\partial\psi}{\partial{y}} \label{vzpsi},
\end{equation}
\begin{equation}
\rho'\kappa\Gamma{v_y} = -\frac{\partial\psi}{\partial{z}} \label{vypsi},
\end{equation}
which manifestly solves the continuity equation; we introduced a factor of $\kappa$, the opacity, to ensure that $\psi$ is dimensionless. These relations also demonstrate that $v_y \sim \delta{v_z}$, which is what we expect: the velocity in the direction perpendicular to the majority of the motion is reduced by a factor of $\delta$. Note, however, that the definition of the stream function now involves the density, which is not constant in this analysis. 

As is done in the standard Blasius treatment, we assume that the stream function varies self-similarly as

\begin{equation}
\psi = \sqrt{\frac{8}{27}\rho'_0\,\kappa\,\Gamma_j\,v_j}\,z^{1/2}f(\alpha) \label{psiselfsim},
\end{equation}
where 

\begin{equation}
\alpha = y/\delta{y} = \frac{y}{\sqrt{z}}\sqrt{\frac{27}{8}\rho'_0\,\kappa\,\Gamma_j\,v_j} \label{alphaeq}
\end{equation}
is our self-similar variable (we used equation \eqref{deltaeq} for the boundary layer thickness) and $f$ is a function to be determined from equations \eqref{zmom1} and \eqref{gasen1}. We will also assume that the density varies self-similarly as

\begin{equation}
\rho' = \rho'_0\,g(\alpha) \label{rhoselfsim},
\end{equation}
where $g$ is a second function. 

One can use these definitions in equations \eqref{zmom1} and \eqref{gasen1} to derive a set of coupled, nonlinear, ordinary differential equations for $f$ and $g$. It is mathematically convenient, however, to define a new self-similar variable $\xi$ by

\begin{equation}
\xi = \int_0^{\alpha}g(\tilde{\alpha})\,d\tilde{\alpha} \label{xiofalpha},
\end{equation}
where $\tilde{\alpha}$ is a dummy variable of integration, and write the functions $f$ and $g$ in terms of this variable. This approach is similar to the one taken by \citet{ara92}.

With this parametrization, the velocities are

\begin{equation}
\Gamma{v_z} = \Gamma_jv_jf_{\xi} \label{vzselfsim},
\end{equation}
\begin{equation}
\rho'\Gamma{v_y} = \frac{1}{2\sqrt{z}}\sqrt{\frac{8}{27}\frac{\rho'_0\,\Gamma_j\,v_j}{\kappa}}\bigg{(}\alpha{f_{\xi}}g-f\bigg{)} \label{vyselfsim},
\end{equation}
where {}{a function with a subscript $\xi$ denotes the derivative of that function with respect to $\xi$, i.e., $f_{\xi} = df/d\xi$, $f_{\xi\xi} = d^2f/d\xi^2$, etc.} Substituting these relations and the self-similar scaling for $\rho'$, equation \eqref{rhoselfsim}, into equations \eqref{zmom1} and \eqref{gasen1}, we find the following self-similar equations for $f$ and $g$:

\begin{equation}
-\frac{1}{2}\bigg{(}g+\frac{4}{3}\mu\bigg{)}f\,f_{\xi\xi}+\mu\,\Gamma_j^2v_j^2\frac{g\,f_{\xi}(f_{\xi\xi})^2}{1+v_j^2\Gamma_j^2(f_{\xi})^2} = \mu\,g\,f_{\xi\xi\xi} \label{selfsim1},
\end{equation}
\begin{equation}
g_{\xi}f = \frac{3}{2}\Gamma_j^2v_j^2\frac{g^2(f_{\xi\xi})^2}{1+\Gamma_j^2v_j^2(f_{\xi})^2} \label{selfsim2},
\end{equation}
where we have defined $\mu$ as the ratio $e'_0/\rho'_0$. The term $-2\,\mu\,f\,f_{\xi\xi}/3$ and the last term on the left-hand side of equation \eqref{selfsim1} were absent in the treatment of \citet{ara92} because they did not include the enthalpy of the radiation. In the non-relativistic, $\mu \ll 1$ limit, equation \eqref{selfsim1} reduces to the standard Blasius equation (by rescaling the self-similar variable) and $g$ and $f$ decouple from one another, as was found by \citet{ara92}. 

Before proceeding further, recall that the two-stream problem separates the jet and ambient material into two distinct media, the dividing line for our problem chosen to be $y = 0$, and that the asymptotic densities attained in these two media are $\rho'_j$ and $\rho'_a$ for the jet and ambient material, respectively. Therefore, in equations \eqref{psiselfsim}, \eqref{alphaeq} and \eqref{rhoselfsim}, the parameter $\rho'_0$ refers to either $\rho'_j$ or $\rho'_a$ depending on the sign of $\alpha$ and, hence, $\xi$. The functions $f$ and $g$ are thus piecewise defined about $y = 0$, as are the self-similar variables $\alpha$ and $\xi$, with solutions for $\xi > 0$ corresponding to jet quantities and those for $\xi < 0$ corresponding to the ambient medium. Equations \eqref{selfsim1} and \eqref{selfsim2} should therefore be interpreted as a total of four equations for four functions: $f$ defined in the jet, $f$ defined in the ambient medium, $g$ defined in the jet, and $g$ defined in the ambient medium. Note that $\mu$ can also differ between the two media, depending on the asymptotic density (but $e'_0$ must be continuous across $y = 0$ because of equation \eqref{ymom1}).

Keeping in mind this subtlety of equations \eqref{selfsim1} and \eqref{selfsim2}, we must additionally determine the boundary conditions satisfied by $f$ and $g$. The first two conditions satisfied by $f$ can be determined by recalling that $v \rightarrow v_j$ as $y \rightarrow \infty$ and $v \rightarrow 0$ as $y \rightarrow -\infty$. Investigating equation \eqref{vzselfsim} and noting that $\xi$ scales with $y$, these requirements translate to

\begin{equation}
f_{\xi}(\infty) = 1, \quad f_{\xi}( -\infty) = 0.
\end{equation}
We also require that the density approach its asymptotic values in the limits of $y \rightarrow \pm \infty$. From equation \eqref{rhoselfsim}, this gives

\begin{equation}
g(\pm\infty) = 1.
\end{equation}

Now, note that if the jet and ambient materials are to retain their respective identities, then the flow along the surface of contact at $y = 0$ must remain parallel to that surface. In other words, there must not be any mass flow across the boundary, i.e., this surface is a contact discontinuity, which means that $v_y(0) = 0$. From equation \eqref{vyselfsim}, this shows that $f$ must satisfy

\begin{equation}
f(0) = 0.
\end{equation}

The other two boundary conditions can be determined by requiring that the normal and transverse components of the energy-momentum tensor, the divergence of which we took to obtain equation \eqref{radhydroco}, be continuous across the point of contact $y = 0$. We can show that these restrictions demand that $f_{\xi}$ and $f_{\xi\xi}$ be continuous across $y = 0$, which closes the system. 

We can also calculate the comoving-frame number density of photons $N'$ by solving equation \eqref{fluxeq}. Keeping only terms to lowest order in the boundary layer thickness, the equation of conservation of photon number becomes

\begin{equation}
\nabla_{\mu}F^{\mu} = \nabla_{\mu}\bigg{(}N'U^{\mu}\bigg{)}-\frac{1}{3}\frac{\partial}{\partial{y}}\bigg{(}\frac{1}{\rho'\kappa}\frac{\partial{N'}}{\partial{y}}\bigg{)} = 0.
\end{equation}
As we did for the number density of scatterers, we will assume that the photon number density varies self-similarly across the boundary layer as

\begin{equation}
N' = N'_0\,h(\xi) \label{Nselfsim},
\end{equation}
where $N'_0$ is the asymptotic number density of the jet or the ambient medium. With this form for the number density, we find that the equation for $h$ becomes

\begin{equation}
\frac{3}{2}\Gamma_j^2v_j^2\frac{g\,(f_{\xi\xi})^2}{1+\Gamma_j^2v_j^2(f_{\xi})^2}h-f\,h_{\xi}=\frac{9}{4}g\,h_{\xi\xi}\label{selfsim3}.
\end{equation}
As for the functions $f$ and $g$, $h$ is really piecewise defined across the boundary $y = 0$, and so equation \eqref{selfsim3} is really two equations -- one for the photon number density in the jet, and another for that in the ambient medium. The requirement that the number density of photons asymptotically approach its jet and ambient values gives 

\begin{equation}
h(\pm\infty) = 1.
\end{equation}
The $y$-component of the flux must also be continuous across the boundary, which, when written out, shows that the derivative of $h$ must be continuous across $\xi = 0$, which yields the final two boundary conditions.

Finally, the solutions to equations \eqref{selfsim1}, \eqref{selfsim2}, and \eqref{selfsim3} will be in terms of the parameter $\xi$, which is itself a function of $g$ (equation \eqref{xiofalpha}). We would like the solutions to be in terms of the parameter $\alpha$, which is directly related to the physical coordinates $y$ and $z$ (equation \eqref{alphaeq}). The transformation can be achieved by differentiating equation \eqref{xiofalpha}, rearranging, and integrating to yield

\begin{equation}
\int_0^{\xi}\frac{d\tilde\xi}{g(\tilde\xi)} = \alpha \label{xiofalpha2},
\end{equation}
where $\tilde\xi$ is an integration variable and we set the integration constant to zero because we demand $\xi(\alpha = 0) = 0$. After solving equations \eqref{selfsim1} and \eqref{selfsim2} for $g(\xi)$, we can numerically integrate and solve equation \eqref{xiofalpha2} for $\xi$ in terms of $\alpha$.

\subsection{Solutions}
Equations \eqref{selfsim1}, \eqref{selfsim2} and \eqref{selfsim3}, together with the boundary conditions on $f$, $g$, and $h$, govern the behavior of the velocity, number density of scatterers, and number density of photons throughout the two-stream boundary layer. Here we present and analyze the solutions to those equations as we vary the asymptotic jet Lorentz factor $\Gamma_j$ and the quantity $\mu = e'_0/\rho'_0$. 

One caveat with the definition of $\alpha$ is that it depends on $\Gamma_j$ via equation \eqref{alphaeq}. Therefore, if we plot solutions with different $\Gamma_j$, we must be careful to incorporate this dependence so that the range of physical space we consider for each solution is the same. Because of this fact, in the following figures we will plot our solutions as functions of the variable

\begin{equation}
\tilde\alpha = \frac{\alpha}{\sqrt{\Gamma_jv_j}} = \frac{y}{\sqrt{z}}\sqrt{\frac{27}{8}\rho'_0\kappa}.
\end{equation}

\begin{figure}[htbp] 
   \centering
   \includegraphics[width=3.5in]{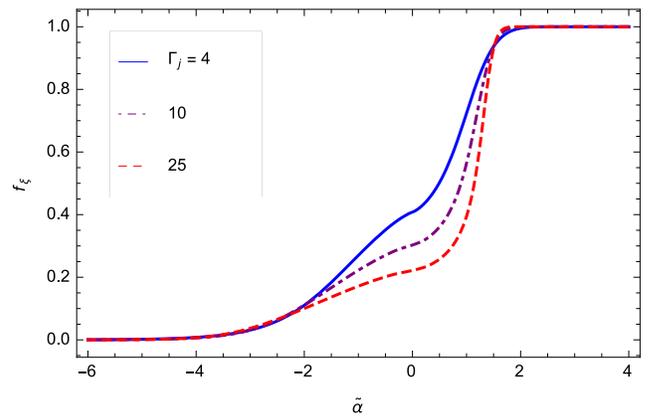} 
   \caption{The function $f_{\xi}$, which is the normalized $z$-component of the four velocity, in terms of the parameter $\tilde\alpha \propto y/\sqrt{z}$, for $\mu = 1$ and a number of jet Lorentz factors, as indicated by the legend. As we can see, the thickness of the velocity boundary layer, in terms of $\tilde\alpha$, is nearly independent of $\Gamma_j$. }
   \label{fig:fpplottsgam}
\end{figure}

\begin{figure}[htbp] 
   \centering
   \includegraphics[width=3.5in]{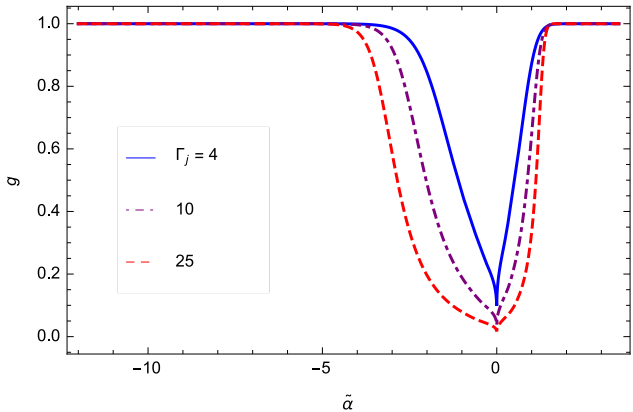} 
   \caption{The variation of the normalized density, given by $g$, as a function of $\tilde\alpha$ for the same parameters as those chosen for Figure \ref{fig:fpplottsgam}. The density remains below its asymptotic value over a slightly larger range of $\tilde\alpha$ for higher $\Gamma_j$, and the decrease in density within the boundary layer is consistent with the findings of \citet{ara92}. {}{The density formally equals zero at $\tilde\alpha = 0$; however, because $g$ approaches zero at a very slow rate (recall $g(\xi) \propto -1/\ln\xi$), it appears from the Figure, which only samples a finite number of points around $\tilde\alpha=0$, that the density remains positive and larger for smaller $\Gamma_j$.} }
   \label{fig:gplottsgam}
\end{figure}

\begin{figure}[htbp] 
   \centering
   \includegraphics[width=3.5in]{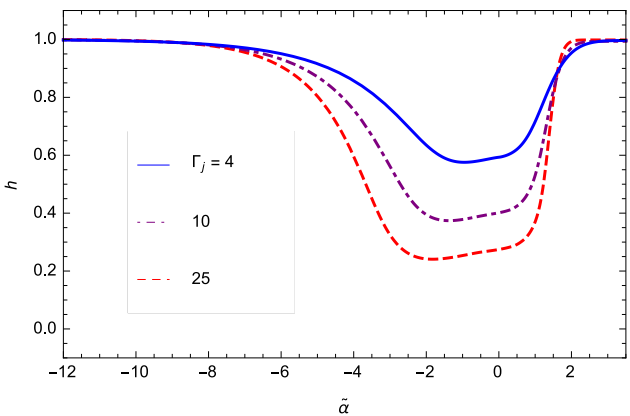} 
   \caption{The solution to equation \eqref{selfsim3}, $h$, which is the normalized number density of photons, for the same set of parameters as those chosen in Figure \ref{fig:fpplottsgam}. The number density of photons is seen to roughly track the number density of scatterers. Because the energy density of the radiation remains constant across the layer, the energy per photon increases in the boundary layer.}
   \label{fig:hplottsgam}
\end{figure}

Figure \ref{fig:fpplottsgam} shows the solution for the normalized $z$-component of the four-velocity ($f_{\xi}$) as a function of the self-similar variable $\tilde\alpha \propto y/\sqrt{z}$ for a number of jet Lorentz factors and $\mu = 1$. We see that the width of the boundary layer is nearly unchanged as we vary the Lorentz factor of the asymptotic jet. The value of $f_{\xi}$ at the contact discontinuity is lower for larger $\Gamma_j$, resulting in a greater shear ($\sim f_{\xi\xi}$) as one proceeds into the jet. The flattening of the velocity around $\tilde\alpha = 0$ arises from the behavior of the density around this region and the function $\xi(\alpha)$ determined therefrom. 

Figure \ref{fig:gplottsgam} illustrates the manner in which the density varies over the boundary layer for the same parameters as those chosen for Figure \ref{fig:fpplottsgam}. Consistent with the findings of \citet{ara92}, we find that the transition from the jet to the ambient medium carves out a region of low density material. This behavior can be understood by noting that the shear in the flow causes viscous heating of the fluid, which results in an increase in the specific entropy $s'$. Since the specific entropy scales as $s' \propto e'/(\rho')^{4/3}$ for a radiation-dominated gas, an increase in the entropy at constant pressure corresponds to an decrease in the density of scatterers. As was found by \citet{ara92}, the density equals zero at the origin, which can be gleaned from the asymptotic behavior of equation \eqref{selfsim2}: for $\xi \ll 1$, we can let $f_{\xi} \sim f_{\xi\xi} \sim \xi$, and solving the resultant approximate differential equation shows that $g \propto -1/\ln(\xi)$. 

In Figure \ref{fig:hplottsgam} we plot the solution to equation \eqref{selfsim3}, the normalized, rest-frame number density of photons, for the same set of Lorentz factors and $\mu = 1$. By comparing this with Figure \ref{fig:gplottsgam}, we see that the photon number density roughly tracks that of the scatterers. However, the photon number density remains below its asymptotic value significantly farther into the ambient medium than does the particle density. Because the energy density of the radiation is constant across the boundary layer, the decrease in photon density corresponds to a higher average energy per photon increases within the boundary layer -- a clear manifestation of viscous heating, as noted by \citet{ara92}.

\begin{figure}[htbp] 
   \centering
   \includegraphics[width=3.5in]{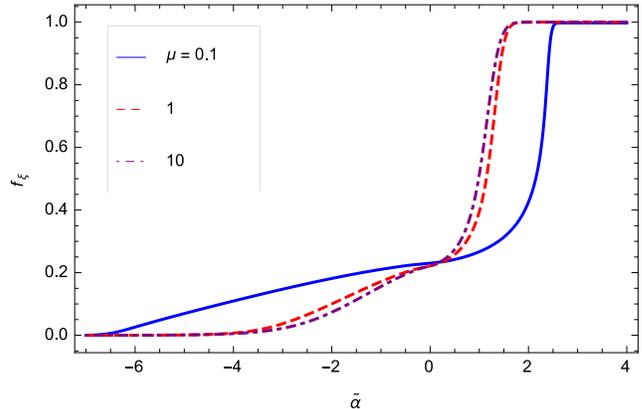} 
   \caption{The solution for the normalized $z$-component of the four-velocity ($f_{\xi}$) for $\Gamma_j = 25$ and three different values of $\mu$, as indicated by the legend. The velocity profile does not differ much as $\mu$ increases beyond 1, but for smaller values of $\mu$ the boundary layer widens noticeably.}
   \label{fig:fpplotstsr}
\end{figure}

\begin{figure}[htbp] 
   \centering
   \includegraphics[width=3.5in]{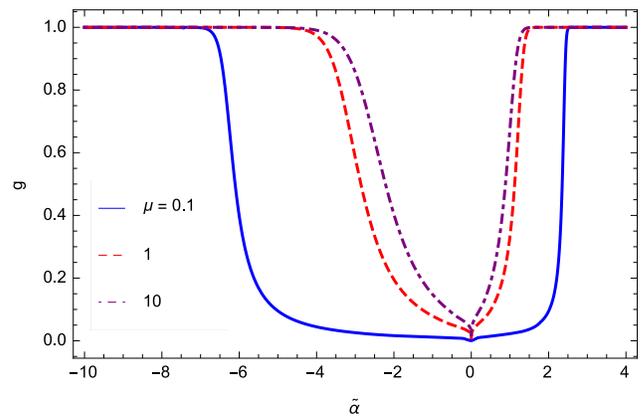} 
   \caption{The function $g$, which is the rest-frame number density of scatterers, for the same parameters as in Figure \ref{fig:fpplotstsr}. For smaller values of $\mu$, the density is significantly reduced from its asymptotic value over a larger range in $\alpha$.}
   \label{fig:gplotstsr}
\end{figure}

\begin{figure}[htbp] 
   \centering
   \includegraphics[width=3.5in]{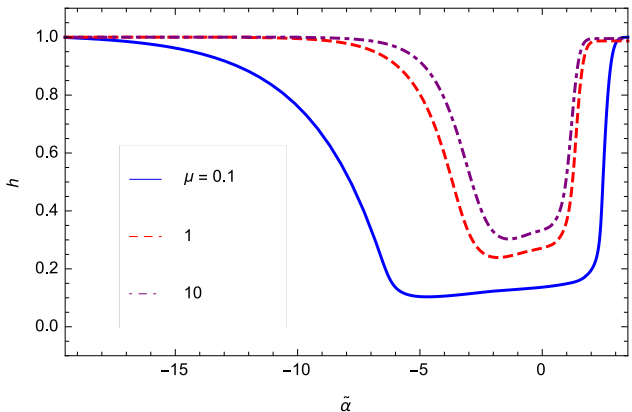} 
   \caption{The solution to equation \eqref{selfsim3}, which gives the number density of photons observed in the comoving frame, for the same set of parameters as in Figure \ref{fig:fpplotstsr}. The radiation number density roughly follows that of the scatterers.}
   \label{fig:hplotstsr}
\end{figure}

Figures \ref{fig:fpplotstsr}, \ref{fig:gplotstsr}, and \ref{fig:hplotstsr} show the $z$-component of the four-velocity, the number density of scatterers, and the photon number density, respectively, for $\Gamma_j = 25$ and $\mu = 0.1$, 1, and 10. {}{The overall qualitative behavior of the solutions is similar to that depicted in Figures \ref{fig:fpplottsgam} -- \ref{fig:hplottsgam}. More specifically, however, we find that values of $\mu$ greater than unity cause} the boundary layer thickness {}{to decrease}, but not appreciably. On the contrary, {}{a value of $\mu$ only marginally less than one results in a significant widening of} the boundary layer. {}{This dependence is ultimately related to the compressibility of the fluid and the relation between that compressibility and the sound speed of a radiation-dominated gas (see discussion below).}

\section{Discussion}
The plots of the previous subsection demonstrate how radiation viscosity affects the boundary between a fast-moving flow and its surroundings under the two-stream approximation. Our analysis generalizes the treatment of \citet{ara92} by permitting the jet velocity to be relativistic and by incorporating the enthalpy of the radiation in the momentum equation. Our results are similar to those found by \citet{ara92} (compare, e.g., their Figure 3 to our Figures \ref{fig:fpplottsgam} and \ref{fig:gplottsgam}); there are, however, a few notable differences.

For one, our boundary layer thickness, given by equation \eqref{deltaeq}, differs from that of \citet{ara92}, who found $\delta^2 \sim e'/(\rho'_0{}^2\kappa{\,}v_jz)$. In addition to the Lorentz factor contained in ours, their boundary layer thickness has an additional factor of $e'/\rho'_0$, meaning that, in the non-relativistic limit, our results do not agree. This discrepancy arises from the fact that, while we compared the lowest-order terms in $\delta$ to the inviscid terms in the gas energy equation to obtain our boundary layer thickness, they compared the viscous term to the inertial term{}{ -- the one} proportional to $\rho'$ {}{--} in the $z$-component of the momentum equation to obtain theirs. Because they ignored the enthalpy of the radiation, the inertial term was the only inviscid term present in the momentum equation, making it the only one available to balance the viscous part. However, if one does not ignore the radiative contribution to the momentum equation, one can now equate the viscous terms to either the term proportional to the mass density or the one proportional to the enthalpy. Because the density equals zero at $\alpha = 0$, there is always some location at which the enthalpy term exceeds the inertial term, making it more reasonable to equate the former to the viscous contribution than the latter. If one follows this route, one recovers our ordering for the boundary layer thickness. 

The second difference is that their solutions depend on the square of the Mach number, $M^2 \equiv v^2/c_s^2$, where $c_s^2 = 4e'/(9\rho'_0)$ is the non-relativistic sound speed. Our solutions, on the other hand, depend on both the jet velocity $v_j$ and the ratio $\mu = e'/\rho'_0$, which is proportional to the non-relativistic sound speed. One reason for this difference arises from the discrepancy between our definitions of the boundary layer thickness. Another is due to the fact that our solutions included the enthalpy of the radiation; had \citet{ara92} included this term, factors of the sound speed would have arisen in their $z$-momentum equation. Finally, our solutions also depend on the value of the jet velocity -- not just the ratio of the jet speed to the sound speed -- because we included all relativistic effects, meaning that the speed of light now plays a role in determining the evolution of the system. 

Our solutions show that, for fixed $\mu$, the thickness of the boundary layer is approximately independent of the jet Lorentz factor, which is due to the competition between the scaling of the fiducial boundary layer thickness $\delta \propto 1/\sqrt{\Gamma_j}$ (see equation \eqref{deltaeq}) and the viscous heating. Specifically, a larger $\Gamma_j$ results in a smaller $\delta$ and a greater shear; this shear increases the specific entropy $s' \propto e'/(\rho')^{4/3}$ and, since the pressure is a constant, this increase in the entropy implies a decrease in the density of scatterers which widens the boundary layer. 

Figures \ref{fig:fpplotstsr}, \ref{fig:gplotstsr}, and \ref{fig:hplotstsr} show how the solutions for the velocity, density of scatterers, and density of photons vary for a fixed jet Lorentz factor (we chose $\Gamma_j = 25$) but for a variable $\mu$. Increasing $\mu$ {}{relative to $\mu = 1$} tends to slightly decrease the boundary layer thickness, while decreasing the value of $\mu$ {}{relative to $\mu=1$} dramatically increases the thickness. This behavior arises from the fact that the viscous heating, which decreases the density of scatterers and widens the boundary layer, is effective when the gas is compressible. The compressibility of the fluid, however, is only important when the flow velocity is supersonic, and we can show that the sound speed of the gas is \citep{cou14b}

\begin{equation}
c_s = \frac{2}{3}\sqrt{\frac{\mu}{1+4\mu/3}},
\end{equation}
which, noting that $e' = 3p'$, where $p'$ is the radiation pressure, correctly reduces to $c_s \simeq \sqrt{4p'/(3\rho')}$ in the limit $\mu \ll 1$ and $c_s \simeq 1/\sqrt{3}$ in the limit $\mu \gg 1$. Thus, when $\mu \ll 1$, the location of the sonic point extends farther into the ambient medium, resulting in a widening of the boundary layer. On the other hand, when $\mu \gg 1$, the sonic point approaches the jet, but only slightly due to the fact that the sound speed approaches a constant as $\mu \rightarrow \infty$. In fact, based on this reasoning, we expect our solutions to be independent of $\mu$ in the large $\mu$ limit, which is indeed reflected in equation \eqref{selfsim1}.

The densities of scatterers and photons both decrease within the boundary layer. This behavior has two interesting consequences. First, the lower number density of scatterers means that the optical depth is lower in the boundary layer. We are therefore able to see farther into the medium along lines of sight that probe regions of high shear. Second, because the energy density of the radiation stays unchanged as we traverse the media, the average energy per photon increases, resulting in a harder spectrum within the boundary layer.

Equation \eqref{ymom1} shows that the radiation energy density, and hence the pressure, is constant across the boundary layer, which is ultimately a statement of the causal connectedness of the jet. This means, equivalently, that the boundary layer thickness $\delta{y}$ can be traversed by a sound wave in less time than it takes the jet material to cross the distance $\delta{z}$. Since the transverse sound speed is related to the isotropic sound speed by $c_{s\perp} = c_s/\Gamma_j$, we find that the boundary layer thickness $\delta$ must satisfy $\delta \lesssim c_s/\Gamma_j$. When this inequality is no longer satisfied, equation \eqref{ymom1} no longer holds, and we must include more terms in all of the boundary layer equations that account for changes in the pressure. 

The solution for $g$ equals zero at $\alpha = 0$ which, as we mentioned, can be determined by inspecting the asymptotic ($\xi \rightarrow 0$) limit of equation \eqref{selfsim2}. This feature was also found by \citet{ara92}, and can be understood physically by noting that, when the jet and the ambient medium initially interacted, the surface separating them served as a discontinuity in velocity, resulting in an infinite shear and entropy generation. Although the solutions presented here illustrate the time-steady state of the system after radiation viscosity has smoothed out the discontinuity, the infinite entropy along the contact discontinuity persists and drives the density to zero. Mathematically, this behavior is ultimately due to the fact that $f(0) = 0$, which itself came from the requirement that there be no mixing at the surface separating the jet and the external environment. This boundary condition is necessary to ensure that the two media retain their respective identities and underlies the two-stream assumption, and it allows us to prescribe different asymptotic properties of those media.

\section{Summary and conclusions}
We applied the relativistic equations of radiation hydrodynamics in the viscous limit to the two-stream boundary layer, expanding on the past work of \citet{ara92}. These equations are applicable as long as changes in the fluid quantities are small over the mean free path of a photon. 

An interesting feature of the solutions presented here is the depression in the number density of scatterers within the boundary layer separating the jet and the ambient medium, which is consistent with the findings of \citet{ara92}. We also showed that the number density of photons $N'$ roughly tracks the density of scatterers, reaching a minimum towards, but not at, the contact discontinuity. Therefore, observers viewing a relativistic outflow with lines of sight that probe regions of high shear see farther into the outflow and they also see a higher energy per photon, and hence a harder spectrum.

Our solutions show that the physical boundary layer thickness does not depend {}{strongly} on the jet Lorentz factor $\Gamma_j$, which results from a competition between viscous heating and the scaling of the fiducial boundary layer thickness $\delta \sim 1/\sqrt{\Gamma_j}$. The dependence of our solutions on $\mu$ arises from the fact that the change in entropy of the flow is related to its compressibility, which is most important where the outflow velocity is supersonic. Since the sound speed scales as $c_s \sim \sqrt{\mu}$ when $\mu \lesssim 1$, the point at which the outflow becomes subsonic extends farther into the ambient medium when $\mu$ is small, resulting in a widening of the boundary layer. When $\mu$ becomes larger than one, however, the sound speed does not increase much, asymptotically approaching $1/\sqrt{3}$, meaning that the sonic point only slightly approaches the jet, yielding a marginal decrease in the boundary layer thickness.

{}{A number of assumptions about the nature of the jet and its surroundings are built in to the two-stream solutions presented here. Specifically, we adopted a two-dimensional, plane-parallel geometry for the flow and its surroundings, and we demanded that there be no pressure gradient ($e'(z) = e'_0$) in the ambient medium. While these assumptions greatly enhanced the tractability of the problem, they somewhat hinder the astrophysical relevance of the solutions, as no systems likely conform precisely to these restrictions. However, in \emph{local} regions of an outflow, where the radius of curvature is large and the pressure can be considered relatively constant, these solutions may be actualized. Therefore, while the physics of an entire jet-disk system may be poorly modeled by the two-stream scenario, local patches of the outflow, where the geometry can be treated as locally flat and the pressure gradient is small, are likely well-described by the solutions presented here.}

The two-stream solutions could be applied to relativistic, radiation dominated jets, such as those that appear during super-Eddington TDEs, the event \emph{Swift} J1644+57 being one such case{}{. The event \emph{Swift} J2058+05, observed shortly after \emph{Swift} J1644, is another source that has been interpreted as a jetted TDE \citep{cen12}. A comparison between the models presented here and these sources could provide valuable information concerning their progenitors and the properties of the jets themselves. The application of these models to long GRBs could likewise prove fruitful, potentially yielding, for example, information concerning the Lorentz factor of the jet, the stellar progenitor, and the direction of the line of sight between the observer and the source.}

{}{These models may also be applicable to jetted X-ray binary systems, or ``microquasars'' \citep{mir99, fen04, fen09}. For example, \citet{ara93} applied their non-relativistic, radiation-viscous solution \citep{ara92} to the source SS 433 \citep{fab04, beg06a}. Since the jets of SS 433 are mildly relativistic, their speeds being $v_j \simeq 0.26\,c$ \citep{mar89}, including the relativistic terms arising from the treatment presented here may place new constraints on the properties of those jets and the surrounding envelope.}

{}{Finally, quasi-stars -- protogalactic gas clouds supported by an accreting black hole -- may also contain jets \citep{beg06b, beg08, cze12}. Since the power radiated by the black hole at the center of a quasi-star supports the overlying gaseous envelope, the mass of the envelope greatly exceeding that of the hole, the black hole accretes supercritically by several orders of magnitude. The jets launched from these systems are therefore radiation-dominated, and as they propagate through the quasi-star envelope, radiation-viscous effects likely dominate the interaction between the two media. Applying the solutions presented here to these systems could then yield information about the properties of these jets and the role they may have played during the epoch of galaxy formation.}

One drawback to these models and their physical application, however, is that they require that the jet and ambient medium be separated by a contact discontinuity, which results in the non-physical vanishing of the density of scatterers at the interface. Furthermore, this prevents the jet from entraining ambient material; while this is not particularly problematic for the two-stream problem, in which the jet is considered infinite in extent, realistic jets have a finite width and total momentum, meaning that the entrainment of ambient material will cause a decrease in the outflow velocity with $z$ that cannot be captured with the two-stream treatment. 

In a companion paper \citep{cou15}, we investigate a different type of viscous boundary layer -- the free-streaming jet model. This model treats the entire system, jet and ambient medium, as a single fluid, which removes the need for a contact discontinuity and allows the density to remain {}{non-zero} throughout the {}{boundary layer}. We also show that the entrainment of ambient material causes an overall slowing of the outflow. 

\acknowledgements
This work was supported in part by NASA Astrophysics Theory Program grant NNX14AB37G, NSF grant AST-1411879, and NASA's Fermi Guest Investigator Program.


\begin{thebibliography}{}
\bibitem[Arav \& Begelman(1992)]{ara92} Arav N., Begelman M.C., 1992, ApJ, 401, 125
\bibitem[Arav \& Begelman(1993)]{ara93} Arav N., Begelman M.C., 1993, ApJ, 413, 700
\bibitem[Begelman et al.(2006a)]{beg06a} Begelman M.C., King A.R., Pringle J.E., 2006a, MNRAS, 370, 399
\bibitem[Begelman et al.(2008)]{beg08} Begelman M.C., Rossi E.M., Armitage P.J., 2008, MNRAS, 387, 1649
\bibitem[Begelman et al.(2006b)]{beg06b} Begelman M.C., Volonteri M., Rees M.J., 2006b, MNRAS, 370, 289
\bibitem[Bersier et al.(2004)]{ber04} Bersier D., Rhoads J., Fruchter A., et al., 2004, GCN, 2544, 1
\bibitem[Blandford et al.(1985)]{bla85} Blandford R.D., Jaroszy\'nski M., Kumar S., 1985, MNRAS, 215, 667
\bibitem[Blandford \& Payne(1982)]{bla82} Blandford R.D., Payne D.G., 1982, MNRAS, 199, 883
\bibitem[Blandford \& Znajek(1977)]{bla77} Blandford R.D., Znajek R.L., 1977, MNRAS, 179, 433
\bibitem[Blasius(1908)]{bla08} Blasius H., 1908, in Tech. Memoranda National Advisory Committee for \\
	Aeronautics (English trans.), 1256
\bibitem[Bloom et al.(2011)]{blo11} Bloom J.S., Giannios D., Metzger B.D., et al., 2011, Sci, 333, 203
\bibitem[Bromberg \& Levinson(2007)]{bro07} Bromberg O., Levinson A., 2007, ApJ, 671, 678
\bibitem[Bromberg et al.(2015)]{bro15} Bromberg O., Granot J., Piran T., 2015, MNRAS, 450, 1077
\bibitem[Burrows et al.(2011)]{bur11} Burrows D.N., Kennea J.A., Ghisellini G., et al., 2011, Nature, 476, 421
\bibitem[Cannizzo et al.(2011)]{can11} Cannizzo J.K., Troja E., Lodato G., 2011, ApJ, 742, 32
\bibitem[Cenko et al.(2012)]{cen12} Cenko S.B., Krimm H.A., Horesh A., et al., 2012, ApJ, 753, 77
\bibitem[Coughlin \& Begelman(2014a)]{cou14a} Coughlin E.R., Begelman M.C., 2014a, ApJ, 781, 82
\bibitem[Coughlin \& Begelman(2014b)]{cou14b} Coughlin E.R., Begelman M.C., 2014b, ApJ, 797, 103
\bibitem[Coughlin \& Begelman(2015)]{cou15} Coughlin E.R., Begelman M.C., 2015, ApJ, accepted
\bibitem[Czerny et al.(2012)]{cze12} Czerny B., Janiuk A., Sikora M., et al., ApJ, 2012, 755, L15
\bibitem[Dado et al.(2008)]{dad08} Dado S., Dar A., De R\'ujula A., 2008, ApJ, 680, 517
\bibitem[Evans \& Kochanek(1989)]{eva89} Evans C.R., Kochanek C.S., 1989, ApJ, 346, L13
\bibitem[Fabrika(2004)]{fab04} Fabrika S., 2004, ASPRv, 12, 1
\bibitem[Fender et al.(2004)]{fen04} Fender R.P., Belloni T.M., Gallo E., 2004, MNRAS, 355, 1105
\bibitem[Fender et al.(2009)]{fen09} Fender R.P., Homan J., Belloni T.M., 2009, MNRAS, 396, 1370
\bibitem[Frail et al.(2001)]{fra01} Frail D.A., Kulkarni S.R., Sari R., et al., 2001, ApJ, 562, L55
\bibitem[Fruchter et al.(1999)]{fru99} Fruchter A.S., Thorsett S.E., Metzger M.R., et al., 1999, ApJ, 519, L13
\bibitem[Fynbo et al.(2006)]{fyn06} Fynbo J.P.U., Watson D., Th\"one C.C., et al., 2006, Nature, 444, 1047
\bibitem[Galama et al.(1998)]{gal98} Galama T.J., Vreeswijk P.M., van Paradijs J., et al., 1998, Nature, 395, 670
\bibitem[Giannios \& Metzger(2011)]{gia11} Giannios D., Metzger B.D., 2011, MNRAS, 416, 2102
\bibitem[Guillochon et al.(2014)]{gui14} Guillochon J., Manukian H., Ramirez-Ruiz E., 2014, ApJ, 783, 23
\bibitem[Kamble et al.(2009)]{kam09} Kamble A., Misra K., Bhattacharya D., et al., 2009, MNRAS, 394, 214
\bibitem[Kelly et al.(2014)]{kel14} Kelley L.Z., Tchekhovskoy A., Narayan R., 2014, MNRAS, 445, 3919
\bibitem[Kochanek(1994)]{koc94} Kochanek C.S., 1994, ApJ, 422, 508
\bibitem[Kohler et al.(2012)]{koh12} Kohler S., Begelman M.C., Beckwith K., 2012, MNRAS, 422, 2282
\bibitem[Kohler \& Begelman(2012)]{koh12b} Kohler S., Begelman M.C., 2012, MNRAS, 426, 595
\bibitem[Kohler \& Begelman(2014)]{koh14} Kohler S., Begelman M.C., 2015, MNRAS, 446,1195
\bibitem[Kundu \& Cohen(2008)]{kun08} Kundu P.K., Cohen I.M., 2008, \emph{Fluid Mechanics}, Academic Press
\bibitem[Lazzati et al.(2007)]{laz07} Lazzati D., Morsony B.J., Begelman M.C., 2007, RSPTA, 365, 1141
\bibitem[Levan et al.(2014)]{lev14} Levan A.J., Tanvir N.R., Fruchter A.S., et al., 2014, ApJ, 792, 115
\bibitem[Loeb \& Laor(1992)]{loe92} Loeb A., Laor A., 1992, ApJ, 384, 115
\bibitem[Loeb \& Ulmer(1997)]{loe97} Loeb A., Ulmer A., 1997, ApJ, 489, 573
\bibitem[L\'opez-C\'amara et al.(2013)]{lop13} L\'opez-C\'amara D., Morsony B.J., Begelman M.C., Lazzati D., 2013, ApJ, 767, 19
\bibitem[MacFadyen \& Woosley(1999)]{mac99} MacFadyen A.I., Woosley S.E., 1999, ApJ, 524, 262
\bibitem[Margon \& Anderson(1989)]{mar89} Margon B., Anderson S. F., 1989, ApJ, 347, 448
\bibitem[Matzner(2003)]{mat03} Matzner C.D., 2003, MNRAS, 345, 575
\bibitem[Mirabel \& Rodr\'iguez(1999)]{mir99} Mirabel I. F., Rodr\'iguez L. F., 1999, ARA\&A, 37, 409
\bibitem[Morsony et al.(2007)]{mor07} Morsony B.J., Lazzati D., Begelman M.C., 2007, ApJ, 665, 569
\bibitem[Mundell et al.(2013)]{mun13} Mundell C.G., Kopac D., Arnold D.M., et al., 2013, Nature, 504, 119
\bibitem[Panaitescu(2007)]{pan07} Panaitescu A., 2007, MNRAS, 380, 374
\bibitem[Racusin et al.(2009)]{rac09} Racusin J.L., Liang E.W., Burrows D.N., et al., 2009, ApJ, 698, 43
\bibitem[Rees(1988)]{ree88} Rees M.J., 1988, Nature, 333, 523
\bibitem[Rees \& M\'esz\'aros(1992)]{ree92} Rees M.J., M\'esz\'aros P., 1992, MNRAS, 258, 41
\bibitem[Strubbe \& Quataert(2009)]{str09} Strubbe L.E., Quataert E., 2009, MNRAS, 400, 2070
\bibitem[Tchekhovskoy et al.(2014)]{tch14} Tchekhovskoy A., Metzger B.D., Giannios D., Kelley L.Z., 2014, MNRAS, 437, 2744
\bibitem[Waxman et al.(1998)]{wax98} Waxman E., Kulkarni S.R., Frail D.A., 1998, ApJ, 497, 288
\bibitem[Weinberg(1971)]{wei71} Weinberg S., 1971, ApJ, 168, 175
\bibitem[Woosley(1993)]{woo93} Woosley S.E., 1993, ApJ, 405, 273
\bibitem[Woosley \& Heger(2006)]{woo06} Woosley S.E., Heger A., 2006, ApJ, 637, 914
\bibitem[Zauderer et al.(2011)]{zau11} Zauderer B.A., Berger E., Soderberg A.M., et al., 2011, Nature, 476, 425
\end{thebibliography}
\end{document}